\documentclass[fleqn,usenatbib]{mnras}

\usepackage{newtxtext,newtxmath}

\usepackage[T1]{fontenc}

\DeclareRobustCommand{\VAN}[3]{#2}
\let\VANthebibliography\thebibliography
\def\thebibliography{\DeclareRobustCommand{\VAN}[3]{##3}\VANthebibliography}

\usepackage{graphicx}	
\usepackage{amsmath}	

\usepackage{color} 
\definecolor{bleudefrance}{rgb}{0.19, 0.55, 0.91}
\definecolor{AmericanRed}{rgb}{0.698, 0.133, 0.204}
\definecolor{AmericanBlue}{rgb}{0.0391, 0.1914, 0.3789}
\definecolor{mustard}{rgb}{.808,.702,.00392}
\definecolor{ketchup}{rgb}{0.781, 0.160, 0.1328}
\definecolor{grey}{rgb}{.5,.5,.5}

\title[Cluster-summed colors]{Intra-cluster Summed Galaxy Colors}

\author[A. R. Nachmann and W. K. Black]{
\href{https://orcid.org/0000-0002-1711-7075}{Alexander R. Nachmann}$^{1}$\thanks{Email: nachmann@umich.edu}, 
\href{https://orcid.org/
0000-0003-4811-7913}{W. K. Black}$^{1,2}$
\\
$^{1}$Department of Physics, University of Michigan, Ann Arbor, MI 48109, USA\\
$^{2}$Leinweber Center for Theoretical Physics, University of Michigan, Ann Arbor, MI 48109, USA 
\vspace{-1cm}
}

\date{} 

\pubyear{2021}

\begin{document}
\label{firstpage}
\pagerange{\pageref{firstpage}--\pageref{lastpage}}
\maketitle

\begin{abstract}
Though cluster-summed luminosities have served as mass proxies, cluster-summed colors have received less attention. 
Since galaxy colors have given useful insights into dust content and specific star formation rates,
this research investigates possible correlations between cluster-summed colors and various observable and intrinsic halo properties for clusters in subsamples of TNG, SDSS, and Buzzard. 

Cluster color--magnitude space shows a peak towards the red and bright corner, drawn there by bright red galaxies. Summing colors across a cluster reduces the scatter in color spaces, since magnitude summing acts somewhat like a weighted average. 
The correlation between these summed colors were $(73 \pm 24)\%$ across all three datasets. 
Summed colors and cluster properties typically had low correlations but ranged up to $\sim 40\%$. 
The correlation between color and mass didn't change significantly with richness threshold for TNG and Buzzard, but for SDSS the correlation decreased dramatically with increasing richness, passing from positive correlation to negative correlation near a richness threshold of ten. 


We also looked at mass proxy scaling relations with richness or magnitude and measured the reduction in mass scatter once we added cluster colors. The reduction was generally insignificant, 
but several large reductions in mass scatter occurred under certain circumstances: high-richness Buzzard mass--magnitude relation saw a reduction of $(19 \pm 28)\%$ while low-richness SDSS saw similar order reductions of $(16 \pm 8)\%$ and $(14 \pm 8)\%$ for the mass--richness and mass--magnitude relations respectively. 
This first look at summed cluster color shows potential in aiding mass proxies under certain circumstances, but more deliberate and thorough investigations are needed to better characterize and make use of cluster-summed colors. 

\end{abstract}



\section{Introduction}
Both star and galaxy colors give useful insights into their respective properties. While star colors give us information about temperature and evolutionary pathways, galaxy colors relate to star formation rate, dust content, and metallicity \citep{Worthey_1994}. 
Just as galaxy color is the summed color of all stars composing a galaxy, we define cluster color as the summed color of all galaxies composing a cluster. 
In this paper, we investigate relationships between cluster colors and other cluster properties with a particular focus on its potential to aid in cluster mass estimation.

Cluster galaxies show several features in color--magnitude space. 
  Optically selected galaxies show a strong dichotomy in colors, composed of a photometrically tight and bright ``red sequence'' and a relatively loose and lackluster ``blue cloud'' \citep{Bower_Lucey_Ellis_1992,Strateva+01,Bell+04}.
  The bright central galaxies of clusters (BCG; brightest cluster galaxy) epitomize the radial trends of clusters, they themselves being usually massive, bright red galaxies. As one moves away from cluster center or moves from heavier halos into lighter halos, galaxies are on average dimmer and bluer \citep{Hansen+09,Varga+21}. 
These features then combine across members of clusters to produce a single cluster color. 

Since red fraction increases with galaxy luminosity, cluster mass, and proximity to cluster core, cluster-summed colors will tend towards red sequence mean color with increasing galaxy count. 
Though small (low galaxy count) clusters will tend to have summed colors near the blue cloud, larger (high galaxy count) clusters will tend to have summed colors closer to red sequence mean color. Cluster-summed color will more likely reside below red sequence mean, dragged down by blue cloud members; essentially no galaxies are redder than the red sequence, so the draw is asymmetrical. 
Thus cluster-summed colors will correlate with mass, such that at a given redshift redder clusters will on average be more massive than bluer clusters. 

In general, scatter of cluster colors will decrease as count of member galaxies $N_{\rm gal}$ increases. 
Summed color acts like a weighted average, with brighter galaxies having more of an effect than dimmer ones. Since larger clusters tend to have many bright red galaxies, cluster color will approach the mean red sequence color, with rapidly falling scatter. 
In contrast, `clusters' with only a couple galaxies will have scatter in color distribution close to that of the full galaxy population.


Inspiration for using cluster colors came in part from considering the cutoffs inherent in current richness definitions. 
The red sequence galaxies near cluster center are generally virialized, which makes them good tracers of the underlying mass \citep{Rozo+09_constraining}. Thus, a count of red galaxies should correlate with cluster mass. 
Richness definition comes with some luminosity and radial cutoff; for example, RedMaPPer only counts galaxies brighter than $0.2 ~ L_*$ within a certain radial extent \citep{Rykoff+14}. 
Though richness definitions are highly dependent on luminosity and radius cutoffs, these cutoffs occur where galaxies tend to be relatively dim. Since summing magnitudes downweights dim objects, cluster colors are less sensitive to these choices. 
This then has the upside of reducing systematic shifts due to faint and questionable members.

\section{Datasets \& Methods}
In this section, we discuss the datasets used as well as their resulting color (and magnitude) distributions. 

We note here that our analysis treats datasets inconsistently (e.g. in their redshift, magnitude, or mass cuts). 
While this restricts our ability to compare results between the three, it does show the effects of different selections, such as highlighting Buzzard's significant running of cluster colors with redshift.

\subsection{Datasets}

\begin{table}\centering
  \begin{tabular}{lccccc}
    \hline 
    Dataset & Type & Redshift range & $M_{\rm cluster}$ \\ 
    \hline 
    TNG & Hydro sim & 0 & $M_{200,c}$ \\ 
    SDSS & Observation & $[0,0.05]$ & $M_{\rm vir}$ \\
    Buzzard & Mock catalog & $[0.05,0.32]$ & $M_{\rm vir}$ \\
    \hline 
  \end{tabular}
  \caption{
    Datasets used in this work
  }
  \label{tab:datasets_used}
\end{table}

We examined three datasets: TNG300-1 \citep[][\href{https://www.tng-project.org/data/docs/specifications/\#sec5k}{Model C}, \href{https://www.tng-project.org/files/TNG300-1_StellarPhot/}{observed frame}] {Nelson+18}, SDSS low-redshift NYU-VAGC \citep{Blanton+05}, and the Buzzard Flock \citep{DeRose+19,DeRose+21}. 

\begin{figure}
  \includegraphics[width=\linewidth] {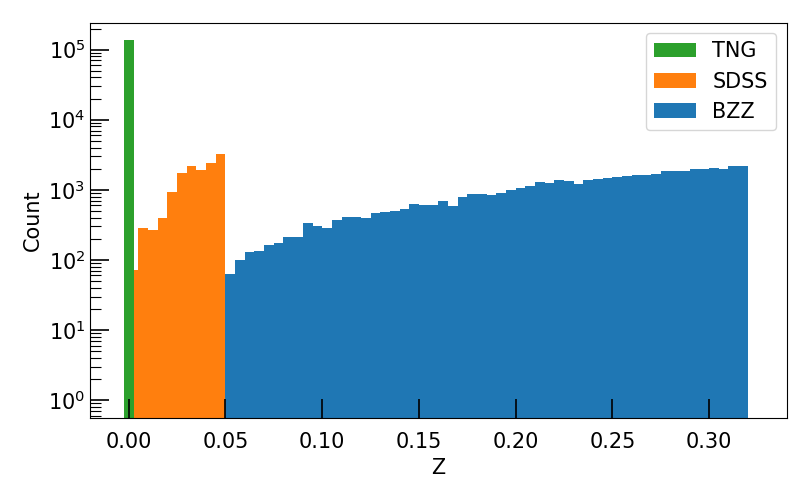}
  \caption{
    Distribution of cluster redshifts in TNG, SDSS, and Buzzard. 
    Note that the TNG data come from redshift slice $Z=0$, hence its appearance as a single bar. 
  }
  \label{fig:histogram_Z}
\end{figure}

\begin{figure}
	\includegraphics[width=\linewidth] {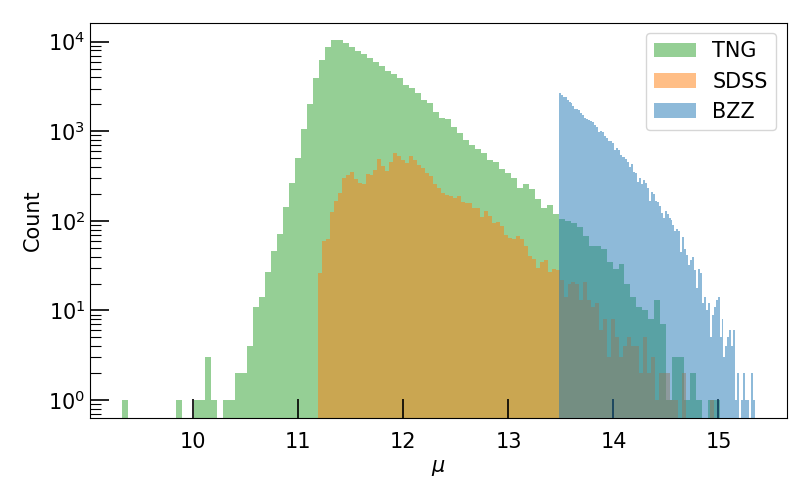}
    \caption{
      Distribution of cluster masses in TNG, SDSS, and Buzzard. 
      The flat left edge in the Buzzard data comes from a $\mu=13.5$ mass cut we implemented. 
      The sharp decline of the left edge in TNG below $\mu=11$ corresponds to stellar suppression \citep{Weinberger+17}. 
      The sharp decline on the left edge of the SDSS data relates to the survey's apparent magnitude limit of eighteenth magnitude. 
    }
    \label{fig:histogram_mu}
\end{figure}

TNG300-1 is a 30~Mpc hydrodynamical simulation---we use the redshift zero slice. Photometric bands were corrected for dust and obstruction that would be present in observed data. No photometric errors were used. 
Clusters in this dataset are defined as all galaxies within $R_{200,c}$ of the central galaxy. We made an $i$ band magnitude cut at $M_i < -18$. 

The SDSS NYU-VAGC is sorted into clusters using a modified version of the Friends of a Friend algorithm by \citet{Lim+17}.  Cluster mass is then calculated as a function of luminosity and magnitude gap \citep{Lu+16}. 
The SDSS spectroscopic data have an apparent magnitude limit of $m_i \lesssim 18$. 
Photometric errors are typically $\gtrsim 0.04~{\rm mag}$ in $griz$. 
We removed clusters beyond $Z=.05$ and excluded photometrically extraneous galaxies. Photometric exclusions used a three-Gaussian model of the galaxies; modeling the red sequence, blue cloud, and photometric outliers (which we excluded). 

The Buzzard Flock dataset (here abbreviated as BZZ) was created using {\sc AddGals}, an algorithm that builds mock galaxy surveys based off of DM-only simulations \citep{Wechsler+21}. The algorithm assigns galaxies to dark matter densities using the luminosity function $\phi(M_r,z)$ so that it matches the luminosity-dependent two point function $P(R_\delta|M_r,z)$. Central galaxies are then added manually at halo centers. 
{\sc AddGals} matches a simulated galaxy with an SDSS one with similar galaxy overdensity and absolute $r$ band magnitude. The SED of the SDSS galaxy is applied to the simulated one and is K-corrected to account for redshift.
This photometry then simulates observational errors. 
Clusters in this dataset are defined as all galaxies within $R_{\rm vir}$ of the central galaxy. Our analysis focused on clusters with $\mu>13.5$ in redshift range $Z|[.05,.32)$. 

Figures~\ref{fig:histogram_Z} and~\ref{fig:histogram_mu} compare distributions in redshift and mass for clusters in each dataset. See appendix~\ref{apx:N_counts} for galaxy count at various thresholds.

\begin{figure*}
	\includegraphics[width=\linewidth]{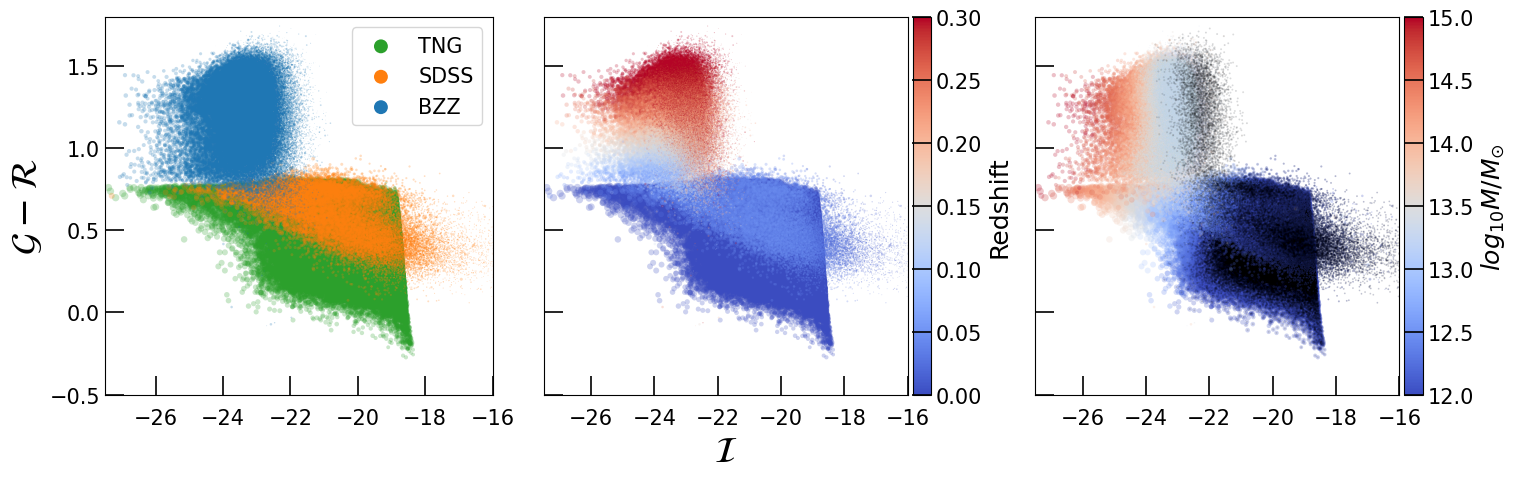}
    \caption{
      Cluster-summed color--magnitude diagrams. 
      Coloring by dataset, redshift, and log mass from left to right. 
      Points sized by mass. 
      In the rightmost panel, single-galaxy `clusters' colored black. 
    }
    \label{fig:CM}
\end{figure*}

\begin{figure*}
	\includegraphics[width=\linewidth]{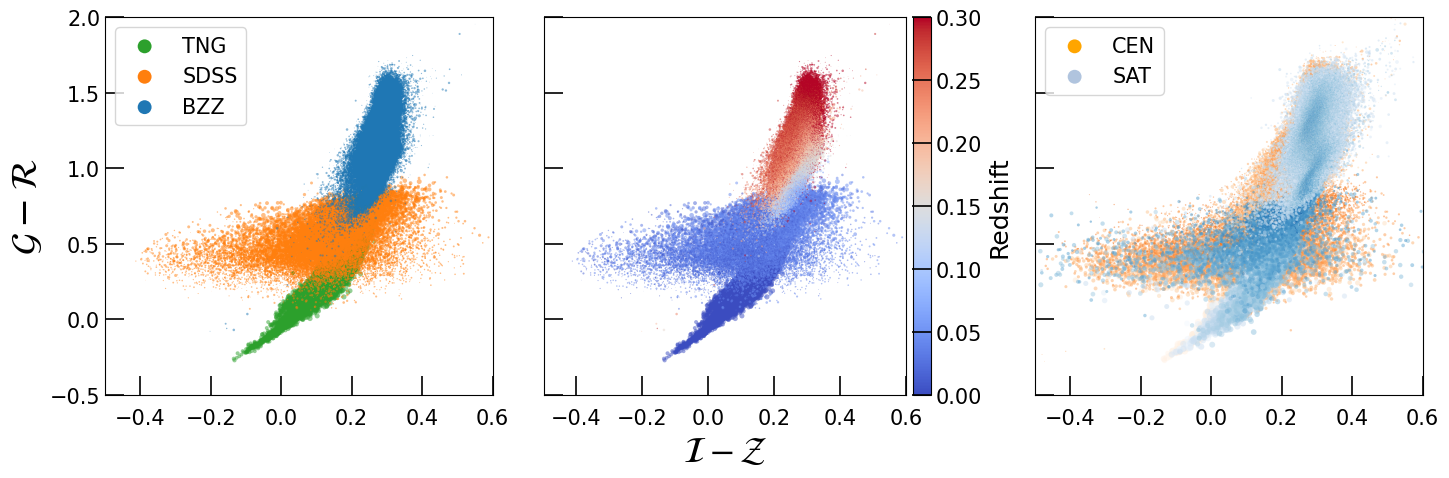}
    \caption{
      Cluster-summed color--color diagram. 
      Coloring by dataset, redshift, and population from left to right. Points sized by mass. 
      In the first two panels, colors are summed across all components (both SAT and CEN), whereas the rightmost panel shows centrals in oranges and summed satellite colors in blue. Darker points in the rightmost panel indicate brighter clusters (in summed $i$ band). 
    }
    \label{fig:CC}
\end{figure*}

\subsection{Definitions}
\label{sec:maths} 
Before summing the magnitudes together, we converted them to absolute magnitudes using the distance modulus (ignoring K 
corrections). Using the low-redshift distance approximation $d = Z\frac{c}{H_0}$ (using $H_0=70~{\rm (km/s)/Mpc}$), we have:
\begin{equation} 
    M_i = m_i-\text{DM} = m_i - 25 -5\log \left( \frac{Z\frac{c}{H_0}} {1~{\rm Mpc}} \right) . 
\end{equation} 
Cluster-summed magnitudes combine fluxes from all member galaxies, then converts the summed flux back into a magnitude. We denote our summed magnitudes with $\mathcal{G}$, $\mathcal{R}$, $\mathcal{I}$, and $\mathcal{Z}$. For example, for $i$ band, we have summed cluster color
\begin{equation}
    \mathcal{I}=\sum M_i = -2.5\log_{10} \left[ \sum_{n=1}^{N} 10^{-\frac{2}{5} (M_i)_n} \right],
\end{equation}
where $N$ is the number of galaxies in the cluster and $(M_i)_n$ is the absolute $i$ band magnitude of the $n$th galaxy. 
In addition to these summed magnitudes, we also compute satellite summed magnitudes (excluding the central galaxy) and investigate relations to BCG magnitudes and colors. We denote these sets as SAT and CEN for satellite-only and central-only respectively. 

Since relations to halo mass generally follow power laws, throughout this paper we define cluster log mass
\begin{equation}
    \mu \equiv \log_{10} M/M_{\odot}
\end{equation}
to simplify equations and graphs. 

Richness for SDSS and TNG is the number of galaxies in a cluster that have $M_i<-21$ and that are members of the redder of the two populations selected by a Gaussian mixture model.
Buzzard richness is calculated using the Red Dragon algorithm \citep{Black+21}, which provides a Gaussian mixture fit continuously evolving across redshift to select the red sequence. To match richness definitions of TNG and SDSS, we used discretized RS probabilities (rounding $P_{\rm red}$ values: $\lambda \equiv \sum \lfloor P_{\rm red} \rceil$).

\subsection{Color--magnitude and color--color diagrams}

Figure~\ref{fig:CM} shows color--magnitude (CM) diagrams for our three data\-sets, displaying several features of note. 
In the top left (bright red) corners of each plot, we see the effect of the red sequence on CM space: summed color is pulled into a peak, shifted up towards red sequence mean color. 
Color distributions show colors of low-mass clusters similarly distributed to the total galaxy population, whereas high-mass cluster colors trend towards the red sequence mean color (at the cluster's respective redshift).

While TNG and SDSS come to form a single point, the redshift span of Buzzard smears out this peak from color $\mathcal{G}-\mathcal{R} = 0.8$ to $1.25$ (a tad slower than the red sequence mean color evolution, which for these low redshifts evolves roughly as $\langle g-r | \, Z \rangle = 0.625 + 3.149 \, Z$; see \citet{Rykoff+12}). 
To a lesser extent, the redness of Buzzard data results from its higher-mass populations; compared to TNG and SDSS, the average Buzzard cluster is far more massive 
(by a factor of $\sim 60$; see figure~\ref{fig:histogram_mu}),
and thus will have a brighter and redder population. 

Figure~\ref{fig:CC} shows color--color (CC) plots. 
Most strikingly, the Buzzard and TNG datasets seem to follow the same curve in CC space whereas SDSS follows a shallower angle. 
The third panel shows the differences between satellite and central populations.
While central colors (CEN) come from lone galaxies, satellite colors (SAT) can be summed over many galaxies ($\sim 6\%$ of TNG and SDSS clusters have more than one galaxy, but this is true for $\sim 80\%$ of Buzzard clusters; see appendix~\ref{apx:N_counts}). Thus as $N$ increases, scatter in SAT trends downwards. 
%



Compared to galaxy CC diagrams, cluster colors have a less distinct red sequence vs blue cloud population. Each of ALL / SAT / CEN tend towards the mean red sequence main color with increasing galaxy count (whereas low-$N$ clusters better mimic the total galaxy population, scattering down towards the blue cloud). 


\section{Results}

\subsection{Trends in color scatter}

\begin{figure}\centering
  \includegraphics [width=\linewidth] {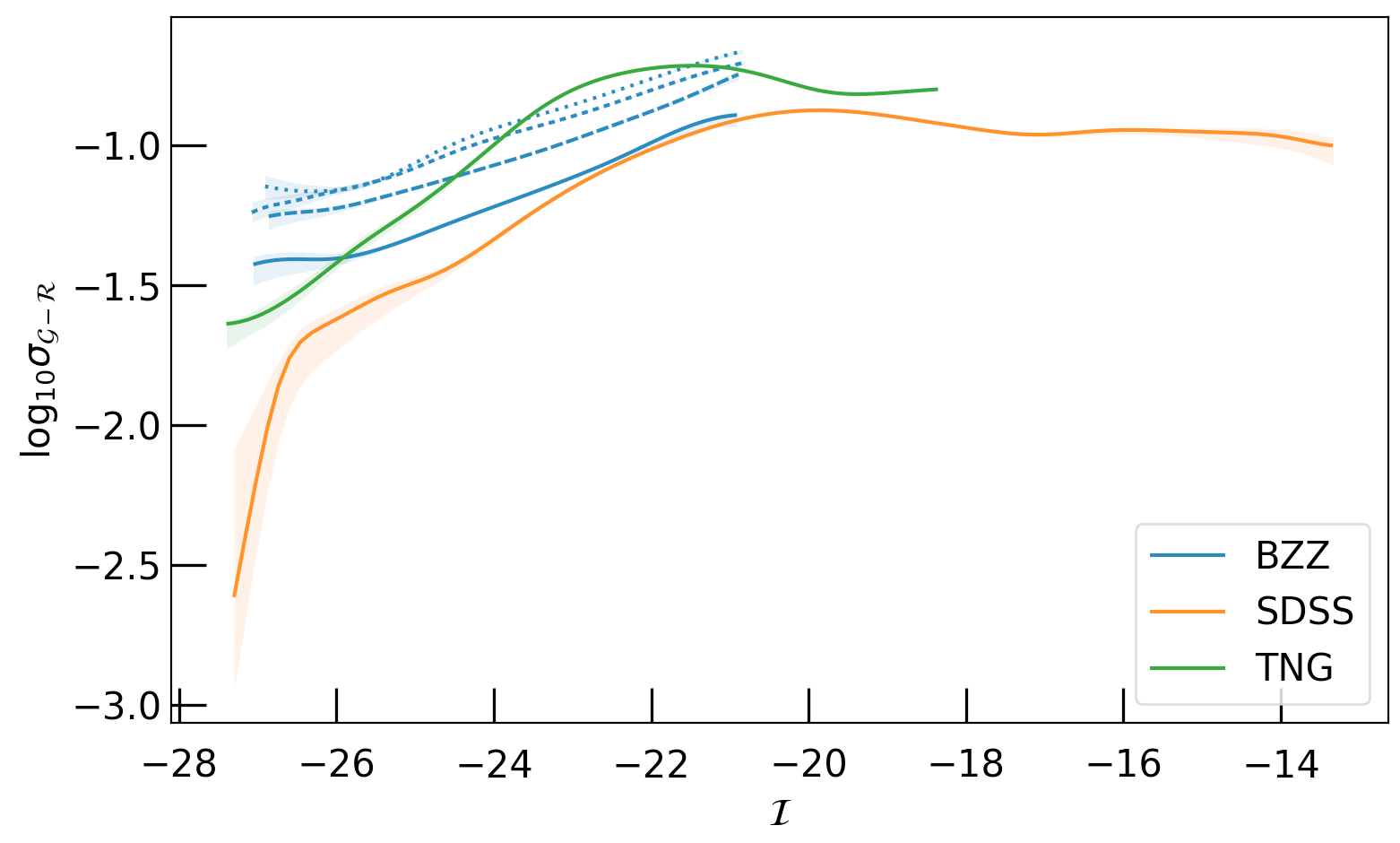}
  \caption{
    Log scatter in $\mathcal{G-R}$ as a function of cluster-summed $i$ band $\mathcal{I}$ with bootstrapped error as transparencies. Buzzard clusters were split into four equal-redshift bins, with higher redshifts more dotted than dashed. Fits performed by 
    \href{https://github.com/afarahi/kllr/tree/master/kllr}{KLLR} 
    \citep{Farahi+18,Anbajagane+20}     
    with Gaussian kernel width of a half magnitude. 
  }
  \label{fig:scatter_vs_iband}
\end{figure}

Figure~\ref{fig:scatter_vs_iband} shows remarkable similarity between simulations. All three datasets feature a slope $\sim 1/6$ increase of log scatter with magnitude until $\mathcal{I} \approx -21$, at which point SDSS and TNG suddenly flatline. Past this point, objects of $\mathcal{I} \gtrsim -21$ behave like randomly-selected galaxies. 
This transition point in scatter behavior could serve as a separating definition between cluster and group scales, where groups lack the scatter reduction trends of clusters.

\subsection{Correlations between colors}

To give a general picture of how cluster-summed colors related to each other and to other observables, we measured covariances between several cluster properties. 

\begin{figure*}
	\includegraphics[width=\linewidth]{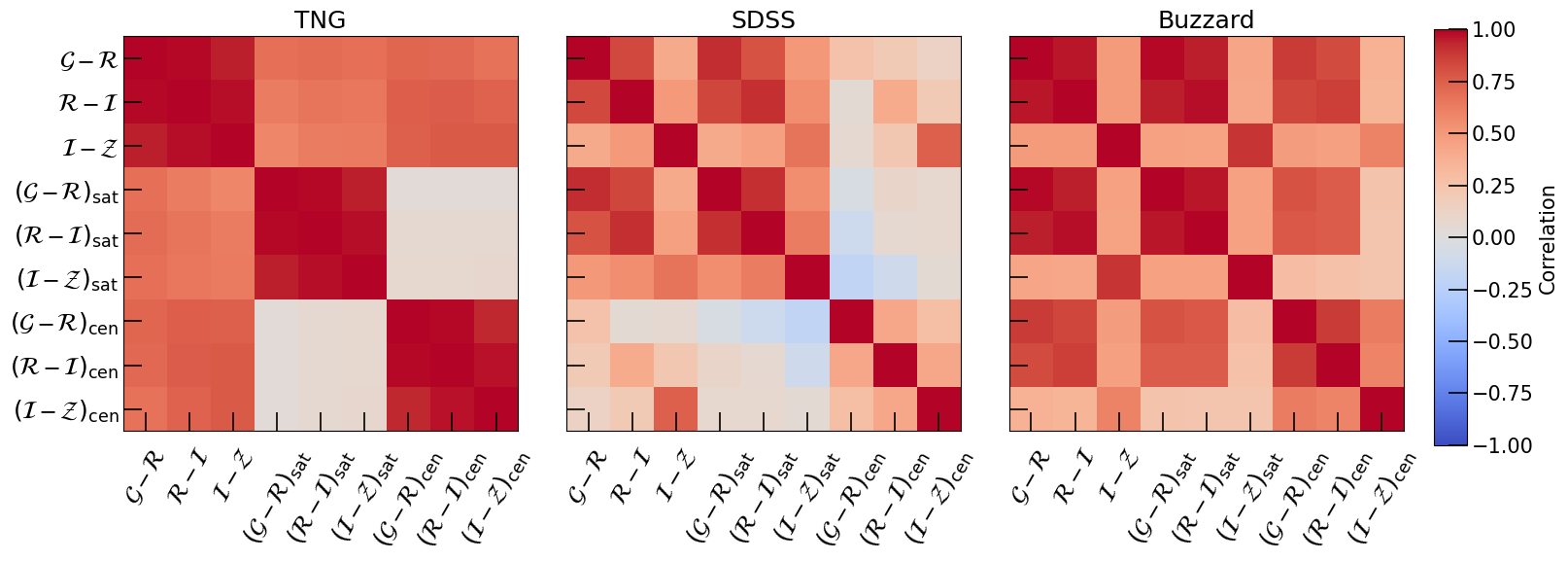}
    \caption{
      Correlation matrix between cluster-summed colors, for each population of ALL, SAT, and CEN. Sample limited to $\lambda > 10$ clusters. 
    }
    \label{fig:corr_matr_colors}
\end{figure*}

Figure~\ref{fig:corr_matr_colors} shows correlations between cluster-summed colors, both between different band combinations ($\mathcal{G-R}$, $\mathcal{R-I}$, $\mathcal{I-Z}$) and populations (ALL / SAT / CEN). Table~\ref{tab:color_correlations} lists summary values for these correlations, grouped by similar behaviors. 

\begin{table}\centering
  \begin{tabular}{l|l}
  \hline 
  Selection & Correlation \\
  \hline
  Different colors, same population & $\rho_{\rm TNG} = 0.97 \pm .02$ \\
  (e.g. $\mathcal{G-R}_{\rm CEN}$ and $\mathcal{R-I}_{\rm CEN}$) & $\rho_{\rm SDSS + BZZ} = 0.61 \pm 0.25$ \\
  \hline
  Identical colors, different populations & $\rho_{\rm BZZ} = 0.78\pm .22$ \\
  (e.g. $\mathcal{G-R}_{\rm CEN}$ and $\mathcal{G-R}_{\rm SAT}$) & $\rho_{\rm SDSS + TNG} = 0.47\pm .33$ \\
  \hline
  Different colors, different populations & $\rho_{\rm SAT \times CEN} = 0.18 \pm 0.22$ \\
  (e.g. $\mathcal{G-R}_{\rm CEN}$ and $\mathcal{R-I}_{\rm SAT}$) & $\rho_{
    \substack{\rm ALL \times SAT \\ \rm \hspace{-.25em} + ALL \times CEN \, }
  } = 0.63 \pm 0.17$ \\
  \hline 
  Any two color definitions & $\rho = 0.53 \pm 0.33$ \\
  \hline 
  \end{tabular}
  \caption{
    Correlations between summed colors for different combinations of population (ALL, SAT, CEN) and color ($\mathcal{G-R}$, $\mathcal{R-I}$, $\mathcal{I-Z}$).
    Some simulations or populations had such similar distributions that we grouped them together (e.g. $\rho_{\rm SDSS+BZZ}$) to highlight notable behavior trends. 
  }
  \label{tab:color_correlations}
\end{table}

We see that for TNG, different cluster colors of the same population type had very strong correlations, typically $\sim 97\%$. 
Buzzard data feature high correlations between identical colors from differing populations, typically $\sim 78\%$. 
In contrast, SDSS cluster colors by any definition were only $\sim 50\%$. 

Satellite and central colors were least correlated, typically only $\sim 18\%$, with SDSS having negative correlations up to $-19\%$. 
Since central color has no consistent correlation with red fraction in our datasets 
(correlations were $24\%$, $-21\%$, and $7.2\%$ for TNG, SDSS, and Buzzard respectively), 
we see no strong evidence that CEN affects SAT. 
We note that---especially for SDSS---raising the richness threshold on the covariance matrix decreased correlations between satellite and central populations, as expected.

We focus the rest of our analysis on the cluster-summed color $(\mathcal{G}-\mathcal{R})$.
At low redshifts, galaxy quenched status (red sequence membership) is better measured with $(g-r)$, so we focus around this distinguishing feature in color space.

\subsection{Correlations with other cluster properties}

\begin{figure*}
	\includegraphics [width=\linewidth] {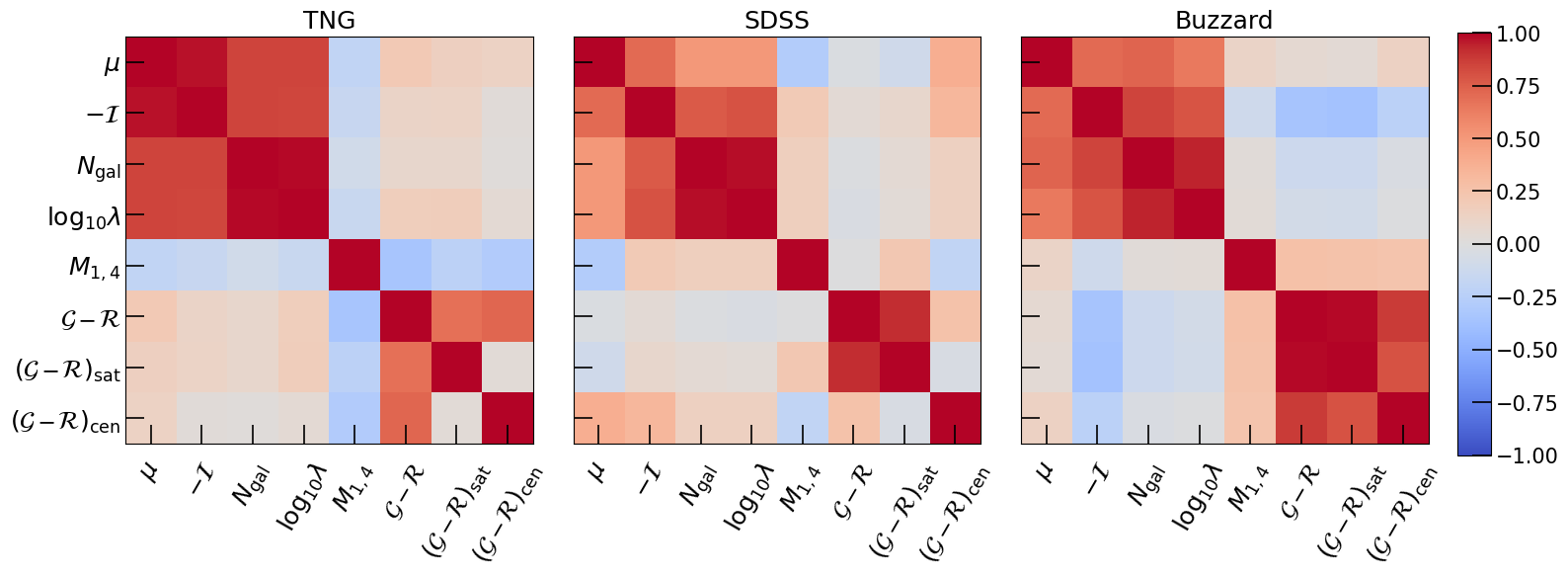} 
    \caption{
      Correlation matrix between cluster-summed colors and other features of the cluster. Sample limited to $\lambda > 10$ clusters. 
    }
    \label{fig:corr_matr_properties}
\end{figure*}

Figure~\ref{fig:corr_matr_properties} shows correlations between cluster $\mathcal{G-R}$ colors and other cluster properties. 
These correlations are generally low, centered around $(4.31\pm 2.8)\%$. 
The few strongest correlations are  
  SDSS $(\mathcal{G-R})_{\rm CEN}$ correlating 39\% with $\mu$ and 
  Buzzard $\mathcal{G-R}$ correlating -35\% with total cluster magnitude $\mathcal{I}$.
We note that $\mathcal{I}$ generally correlated most with log cluster mass $\mu$, followed by galaxy count $N_{\rm gal}$, and richness $\lambda$. 
This hints that summed magnitude may serve as a better mass proxy than richness. 
It also hints that information from blue cloud galaxies can aid in cluster mass estimation.

\subsection{Running of correlations with various thresholds}

The correlation matrices of figures~\ref{fig:corr_matr_colors} and~\ref{fig:corr_matr_properties} give limited information, since correlations vary across different thresholds (such as for high vs low richness clusters). This section investigates how several correlations evolve with richness or mass threshold.

\begin{figure*}
  \includegraphics[width=\linewidth]{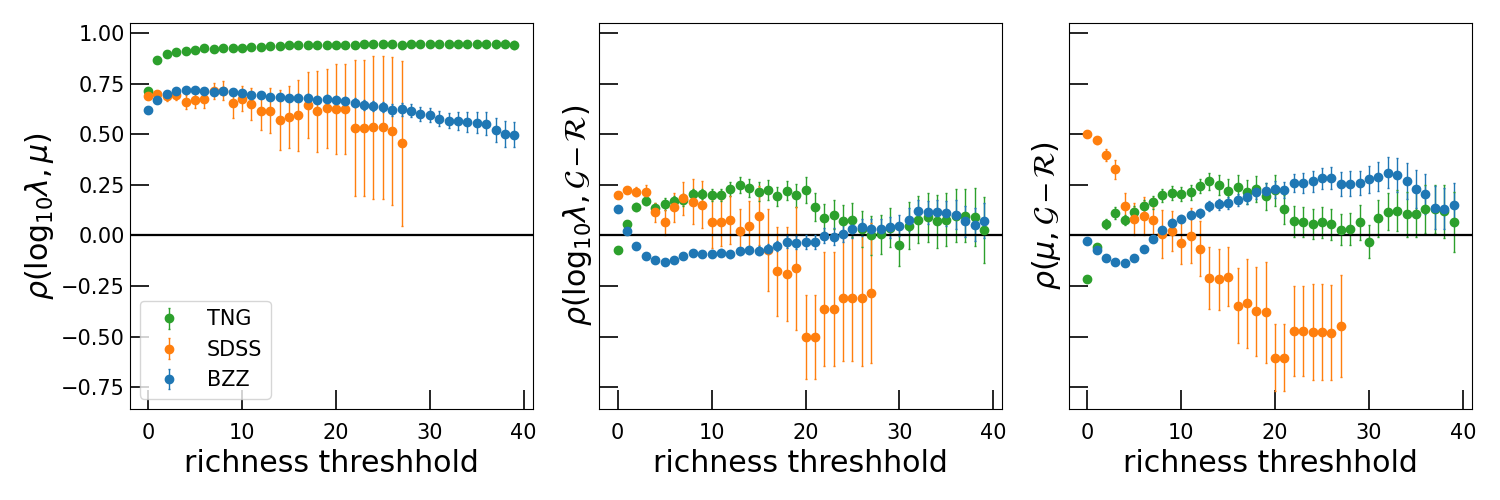}
  \caption{
    Correlation between various cluster properties as a function of richness threshold. From left to right, correlations are measured between cluster mass and richness, between richness and color, and between mass and color. 
    Points only shown for thresholds with $N_{\rm clusters}>10$. 
  }
  \label{fig:rho(richness)}
\end{figure*}

\begin{figure*}
	\includegraphics[width=\linewidth]{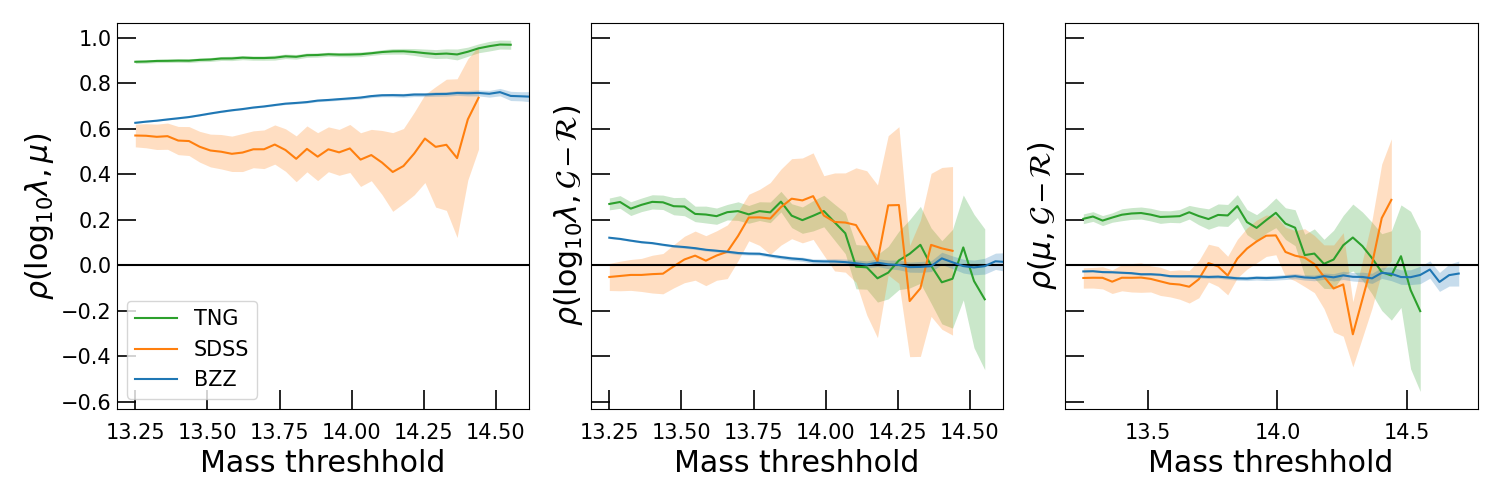}
    \caption{
      Correlation between various cluster properties as a function of mass threshold. From left to right, correlations are measured between cluster mass and richness, between richness and color, and between mass and color. 
    }
    \label{fig:rho(mass)}
\end{figure*}

Figures~\ref{fig:rho(richness)} and~\ref{fig:rho(mass)} show that property correlations have significant richness and mass dependence, though the trends were not always consistent across datasets. 
We found a relatively constant and generally high ($75\% \pm 8\%$) correlation between mass and richness, as expected.

Interestingly, for SDSS the correlations between color and mass $\rho(\mu, \mathcal{G-R})$ and color and richness $\rho(\log_{10} \lambda, \mathcal{G-R})$ both decrease with increasing richness thresholds.
Near a richness threshold of ten, the property correlations with color move from positive to negative. 
In contrast, these correlations show little evolution with mass threshold.

\subsection{Significance of mass scaling}

Using \texttt{emcee} \citep{emcee}, we fit log mass scaling relations
\begin{equation} \label{eqn:mass(lambda)}
  \mu = \alpha \, \log_{10} \lambda +  \beta \,  (\mathcal{G}-\mathcal{R}) + \gamma \, \log_{10} \lambda (\mathcal{G}-\mathcal{R}) \pm \sigma  
\end{equation}
and
\begin{equation} \label{eqn:mass(I)}
  \mu = \alpha \, \mathcal{I} +  \beta \,  (\mathcal{G}-\mathcal{R}) + \gamma \, \mathcal{I}  (\mathcal{G}-\mathcal{R}) \pm \sigma 
\end{equation}
in order to quantify the reduction of log mass scatter $\sigma$ upon adding summed color into the relations. 
After initially setting $\beta = \gamma = 0$ (in order to measure $\sigma$ without the added information from colors), we allowed the parameters to freely vary, after which we measured reduction in scatter. 

\begin{table*}\centering
  \begin{tabular}{l r r l l l l l l}
    \hline 
    & ${\lambda_{\rm min}}$ & ${N_{\rm cluster}}$ & ${\sigma_{[\log_{10}\lambda]}}$ & $\sigma_{[\log_{10}\lambda \, \oplus \, (\mathcal{G}-\mathcal{R})]}$ &
    \% reduction &
    $\sigma_{[\mathcal{I}]}$ &
    $\sigma_{[\mathcal{I} \, \oplus \, (\mathcal{G}-\mathcal{R})]}$&
    \% reduction \\ 
    \hline 
    TNG & 0 & 23 931 
      &  $0.342 \pm 0.002$
      &  $0.356 \pm 0.002$ 
      & $-4.14\pm 0.7$ 
      &  $0.155 \pm 0.0007$  
      &  $0.153 \pm 0.0007$ 
      &  $0.711\pm 0.64$\\
    & 20 & 65 
      &  $0.079 \pm 0.004$   
      &  $0.076 \pm 0.004$
      &  $4.17\pm 7.8$
      &  $0.042 \pm 0.003$   
      &  $0.049 \pm 0.003$ 
      &$-16.74\pm 10.74$\\ 
    SDSS & 0  & 11 844 
      &  $0.326 \pm 0.02$ 
      &  $0.280 \pm 0.02$ 
      & $14.09\pm 8.1$
      &  $0.178 \pm 0.01$ 
      &  $0.150 \pm 0.01$
      & $15.64\pm 8.0$\\ 
    & 20 & 19     
      &  $0.21 \pm 0.43$ 
      &  $0.22 \pm 0.49$
      & $-2.26\pm 310.6$
      &  $0.22 \pm 0.45$ 
      &  $0.23 \pm 0.50$ 
      & $-1.68\pm 305.9$\\ 
    Buzzard  & 0  & 115 508 
      &  $0.2401 \pm 0.005$ 
      &  $0.2379 \pm 0.006$ 
      &  $0.949\pm 3.3$
      &  $0.2190 \pm 0.004$ 
      &  $0.2187 \pm 0.005$ 
      &  $0.123\pm 3.4$\\ 
    & 20 & 865     
      &  $0.164 \pm 0.04$ 
      &  $0.154 \pm 0.04$ 
      &  $6.15\pm 32.0$
      &  $0.174 \pm 0.04$ 
      &  $0.141 \pm 0.03$ 
      & $18.89\pm 28.1$\\
    \hline 
  \end{tabular}
  \caption{
    Scatter in scaling relations of equations~\ref{eqn:mass(lambda)} and~\ref{eqn:mass(I)}. 
    The columns with the first and third scatters correspond to models with null scaling with color (i.e. where $\beta = \gamma = 0$). 
    The columns with the second and fourth scatters correspond to models which permit running of mass with $(\mathcal{G-R})$ color, so comparing odd to even scatter columns shows reduction of scatter due to information from colors. 
  }
  \label{tab:MCMC_scatters}
\end{table*}

Table~\ref{tab:MCMC_scatters} shows measured scatters for our three datasets with richness thresholds $\lambda_{\min} = \{0,20\}$. We found that scatter was typically reduced by $(3.2\pm 2.7)\%$ for mass--richness relations and $(2.8 \pm 5.3)\%$ for mass--magnitude relations across all three datasets.  The most notable of these reductions are high-richness Buzzard $\mu(\mathcal{I},\mathcal{G} - \mathcal{R})$ with a reduction of $18.9\%$. Low-richness SDSS saw a reduction of scatter of $15.6\%$ and $14.1\%$ for $\mu(\mathcal{I},\mathcal{G} - \mathcal{R})$ and $\mu(\log_{10}\lambda,\mathcal{G} - \mathcal{R})$ respectively.
We note that TNG and SDSS have relatively low statistical power above $\lambda \sim 20$ due to low cluster counts (see table~\ref{tab:count_thresholds}).

As hinted at by the high correlations between color definitions (i.e. ALL/SAT/CEN; see figure~\ref{fig:corr_matr_colors}), we found no significant change in scatter reduction when using SAT or CEN instead of ALL. On median, ALL performed slightly superior to the CEN and SAT samples, reducing scatter by 3\% more.

\begin{figure}
  \includegraphics [width=\linewidth] {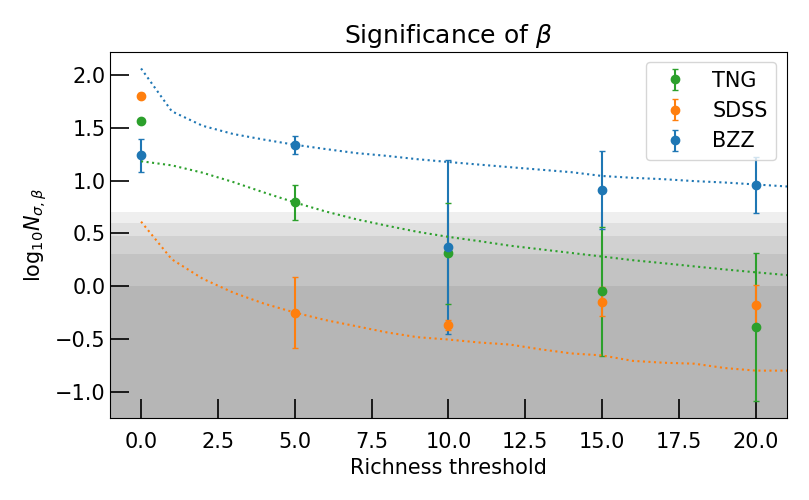}
  \caption{
    Significance of a non-zero $\beta$ parameter (linear correlation between color and mass); points show mean and scatter of the significance values between the richness- and magnitude-based fits of equations~\ref{eqn:mass(I)} and~\ref{eqn:mass(lambda)}.  
    Dotted lines show expected reduction in significance due to Poisson scatter, normalized to the $\lambda > 5$ values. 
    Shaded areas show 1--5$\sigma$ significance levels. 
  }
  \label{fig:beta_sig}
\end{figure}

\begin{figure}
  \includegraphics [width=\linewidth] {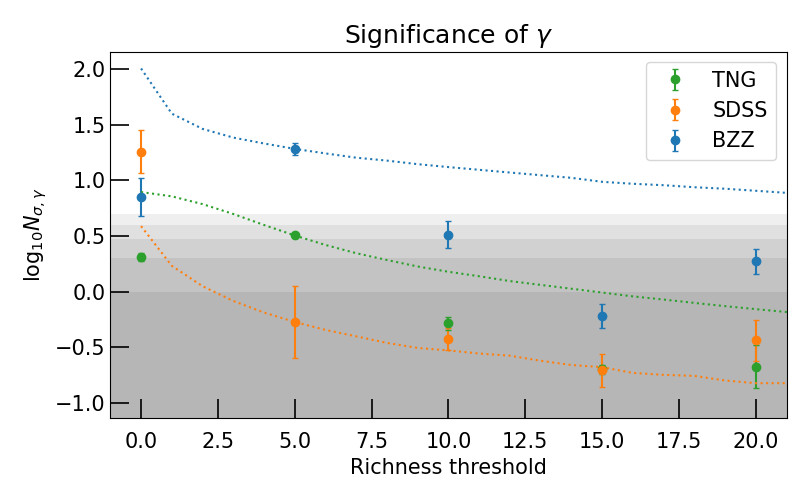}
  \caption{
   As figure~\ref{fig:beta_sig}, but for $\gamma$ (cross term between color and observable). 
  }
  \label{fig:gamma_sig}
\end{figure}

Figures~\ref{fig:beta_sig} and~\ref{fig:gamma_sig} show significance of adding color to equations~\ref{eqn:mass(lambda)}(non-zero $\beta$ and $\gamma$ parameters)

Figures~\ref{fig:beta_sig} and~\ref{fig:gamma_sig} show significance of non-zero $\beta$ and $\gamma$ parameters (of equations~\ref{eqn:mass(lambda)} and~\ref{eqn:mass(I)}). The three datasets behave quite differently, though significance (sigma from zero; $N_\sigma$) generally decays in a Poissonian manner. 
Since Buzzard has more galaxies at most richness thresholds, followed by TNG then SDSS (see figure~\ref{fig:count_vs_threshold}), we see Buzzard showing the most significant running with color.

\section{Conclusions}
In this paper, we investigated cluster-summed colors (i.e. the combined colors of all member galaxies for individual clusters). 
In particular, we looked at their distribution in color--color and color--magnitude space, correlations with cluster properties, how those correlations varied with various thresholds, and how cluster colors impact mass scaling relations.

\begin{itemize}
  \item Cluster-summed color--magnitude diagrams differ from those of galaxies; the summed diagrams have a more continuous distribution which moves towards a bright red peak with mass. 
  \item Cluster-summed colors are weakly correlated with both richness and mass, with strength of correlation varying over richness or mass threshold. Noticeably, the correlation between summed color and mass for SDSS at $\lambda>0$ is about $\sim9\%$, but is $\sim-55\%$ at $\lambda > 20$. 
  \item Cluster-summed magnitude $\mathcal{I}$ may serve as a better mass proxy instead of richness $\lambda$. In all but one dataset and richness threshold, the scatter on the mass relation was smaller using $\mathcal{I}$ than $\log_{10} \lambda$. The only exception is high-richness thresholded SDSS.
\end{itemize}

This study shows that cluster-summed colors can be used to improve mass estimates, especially at low richness thresholds. This was a first-pass analysis to look into possible use of summed colors; their use should be looked into more rigorously in the future.

\section*{Acknowledgements}

We thank Joe DeRose for supplying Buzzard data and Dylan Nelson and Dhayaa Anbajagane supplying TNG data. 
We also thank Eric Bell for his helpful suggestions regarding this project. 

Our work was made possible by the generous open-source software of \href{https://matplotlib.org/} {\sc Matplotlib} \citep{pyplot}, \href{https://numpy.org/doc/stable/} {\sc NumPy} \citep{numpy}, \href{https://scikit-learn.org/stable/} {\sc Sci-kit Learn} \citep{sklearn}, \href{https://docs.h5py.org/en/stable/} {h5py} \citep{h5py}, and \href{https://github.com/afarahi/kllr/tree/master/kllr} {KLLR} \citep{Farahi+18,Anbajagane+20}. 


\section*{Data Availability}
Input datasets:
\begin{itemize}
  \item \href{https://www.tng-project.org/data/docs/specifications/#sec5k}{link} to TNG SDSS-like photometry; data for Model C \href{https://www.tng-project.org/files/TNG300-1_StellarPhot/}{here}
  \item \href{http://sdss.physics.nyu.edu/vagc/lowz.html}{link} to SDSS NYU low-redshift VAGC 
  \item Buzzard Flock available by request \citep[see][]{DeRose+19}
\end{itemize}

\noindent Cluster color datasets calculated in this analysis are all available on Zenodo \citep[see][]{zenodo_repo}.


\bibliographystyle{mnras}
\bibliography{main} 

\begin{thebibliography}{}
\makeatletter
\relax
\def\mn@urlcharsother{\let\do\@makeother \do\$\do\&\do\#\do\^\do\_\do\%\do\~}
\def\mn@doi{\begingroup\mn@urlcharsother \@ifnextchar [ {\mn@doi@}
  {\mn@doi@[]}}
\def\mn@doi@[#1]#2{\def\@tempa{#1}\ifx\@tempa\@empty \href
  {http://dx.doi.org/#2} {doi:#2}\else \href {http://dx.doi.org/#2} {#1}\fi
  \endgroup}
\def\mn@eprint#1#2{\mn@eprint@#1:#2::\@nil}
\def\mn@eprint@arXiv#1{\href {http://arxiv.org/abs/#1} {{\tt arXiv:#1}}}
\def\mn@eprint@dblp#1{\href {http://dblp.uni-trier.de/rec/bibtex/#1.xml}
  {dblp:#1}}
\def\mn@eprint@#1:#2:#3:#4\@nil{\def\@tempa {#1}\def\@tempb {#2}\def\@tempc
  {#3}\ifx \@tempc \@empty \let \@tempc \@tempb \let \@tempb \@tempa \fi \ifx
  \@tempb \@empty \def\@tempb {arXiv}\fi \@ifundefined
  {mn@eprint@\@tempb}{\@tempb:\@tempc}{\expandafter \expandafter \csname
  mn@eprint@\@tempb\endcsname \expandafter{\@tempc}}}

\bibitem[\protect\citeauthoryear{Anbajagane, Evrard, Farahi, Barnes, Dolag,
  McCarthy, Nelson  \& Pillepich}{Anbajagane et~al.}{2020}]{Anbajagane+20}
Anbajagane D.,  Evrard A.~E.,  Farahi A.,  Barnes D.~J.,  Dolag K.,  McCarthy
  I.~G.,  Nelson D.,   Pillepich A.,  2020, \mn@doi [Monthly Notices of the
  Royal Astronomical Society] {10.1093/mnras/staa1147}, 495, 686

\bibitem[\protect\citeauthoryear{Bell et~al.,}{Bell et~al.}{2004}]{Bell+04}
Bell E.~F.,  et~al., 2004, \mn@doi [The Astrophysical Journal]
  {10.1086/420778}, 608, 752

\bibitem[\protect\citeauthoryear{Black \& Evrard}{Black \&
  Evrard}{2021}]{Black+21}
Black W.~K.,  Evrard A.~E.,  2021, in prep.

\bibitem[\protect\citeauthoryear{Blanton et~al.,}{Blanton
  et~al.}{2005}]{Blanton+05}
Blanton M.~R.,  et~al., 2005, \mn@doi [The Astronomical Journal]
  {10.1086/429803}, 129, 2562

\bibitem[\protect\citeauthoryear{Bower, Lucey  \& Ellis}{Bower
  et~al.}{1992}]{Bower_Lucey_Ellis_1992}
Bower R.~G.,  Lucey J.~R.,   Ellis R.~S.,  1992, \mn@doi [Monthly Notices of
  the Royal Astronomical Society] {10.1093/mnras/254.4.601}, 254, 601

\bibitem[\protect\citeauthoryear{Collette et~al.,}{Collette
  et~al.}{2017}]{h5py}
Collette A.,  et~al., 2017, GitHub

\bibitem[\protect\citeauthoryear{DeRose et~al.,}{DeRose
  et~al.}{2019}]{DeRose+19}
DeRose J.,  et~al., 2019, arXiv:1901.02401 [astro-ph]

\bibitem[\protect\citeauthoryear{DeRose et~al.,}{DeRose
  et~al.}{2021}]{DeRose+21}
DeRose J.,  et~al., 2021, arXiv e-prints, p. arXiv:2105.13547

\bibitem[\protect\citeauthoryear{Farahi, Evrard, McCarthy, Barnes  \&
  Kay}{Farahi et~al.}{2018}]{Farahi+18}
Farahi A.,  Evrard A.~E.,  McCarthy I.,  Barnes D.~J.,   Kay S.~T.,  2018,
  \mn@doi [Monthly Notices of the Royal Astronomical Society]
  {10.1093/mnras/sty1179}, 478, 2618

\bibitem[\protect\citeauthoryear{Foreman-Mackey, Hogg, Lang  \&
  Goodman}{Foreman-Mackey et~al.}{2013}]{emcee}
Foreman-Mackey D.,  Hogg D.~W.,  Lang D.,   Goodman J.,  2013, \mn@doi
  [Publications of the Astronomical Society of the Pacific] {10.1086/670067},
  125, 306–312

\bibitem[\protect\citeauthoryear{Hansen, Sheldon, Wechsler  \& Koester}{Hansen
  et~al.}{2009}]{Hansen+09}
Hansen S.~M.,  Sheldon E.~S.,  Wechsler R.~H.,   Koester B.~P.,  2009, \mn@doi
  [The Astrophysical Journal] {10.1088/0004-637X/699/2/1333}, 699, 1333

\bibitem[\protect\citeauthoryear{Hunter}{Hunter}{2007}]{pyplot}
Hunter J.~D.,  2007, \mn@doi [Computing in Science Engineering]
  {10.1109/MCSE.2007.55}, 9, 90

\bibitem[\protect\citeauthoryear{Lim, Mo, Lu, Wang  \& Yang}{Lim
  et~al.}{2017}]{Lim+17}
Lim S.~H.,  Mo H.~J.,  Lu Y.,  Wang H.,   Yang X.,  2017, \mn@doi [Monthly
  Notices of the Royal Astronomical Society] {10.1093/mnras/stx1462}, 470,
  2982–3005

\bibitem[\protect\citeauthoryear{Lu et~al.,}{Lu et~al.}{2016}]{Lu+16}
Lu Y.,  et~al., 2016, \mn@doi [The Astrophysical Journal]
  {10.3847/0004-637X/832/1/39}, 832, 39

\bibitem[\protect\citeauthoryear{Nachmann \& Black}{Nachmann \&
  Black}{2021}]{zenodo_repo}
Nachmann A.~R.,  Black W.~K.,  2021, Inter-cluster summed galaxy colors,
  \mn@doi{10.5281/zenodo.5639341}

\bibitem[\protect\citeauthoryear{Nelson et~al.,}{Nelson
  et~al.}{2018}]{Nelson+18}
Nelson D.,  et~al., 2018, \mn@doi [Monthly Notices of the Royal Astronomical
  Society] {10.1093/mnras/stx3040}, 475, 624

\bibitem[\protect\citeauthoryear{Pedregosa et~al.,}{Pedregosa
  et~al.}{2011}]{sklearn}
Pedregosa F.,  et~al., 2011, the Journal of machine Learning research, 12, 2825

\bibitem[\protect\citeauthoryear{Rozo et~al.,}{Rozo
  et~al.}{2009}]{Rozo+09_constraining}
Rozo E.,  et~al., 2009, \mn@doi [The Astrophysical Journal]
  {10.1088/0004-637X/699/1/768}, 699, 768–781

\bibitem[\protect\citeauthoryear{Rykoff et~al.,}{Rykoff
  et~al.}{2012}]{Rykoff+12}
Rykoff E.~S.,  et~al., 2012, \mn@doi [The Astrophysical Journal]
  {10.1088/0004-637X/746/2/178}, 746, 178

\bibitem[\protect\citeauthoryear{Rykoff et~al.,}{Rykoff
  et~al.}{2014}]{Rykoff+14}
Rykoff E.~S.,  et~al., 2014, \mn@doi [The Astrophysical Journal]
  {10.1088/0004-637X/785/2/104}, 785, 104

\bibitem[\protect\citeauthoryear{Strateva et~al.,}{Strateva
  et~al.}{2001}]{Strateva+01}
Strateva I.,  et~al., 2001, \mn@doi [The Astronomical Journal]
  {10.1086/323301}, 122, 1861

\bibitem[\protect\citeauthoryear{Varga et~al.,}{Varga et~al.}{2021}]{Varga+21}
Varga T.~N.,  et~al., 2021, arXiv:2102.10414 [astro-ph]

\bibitem[\protect\citeauthoryear{Wechsler, DeRose, Busha, Becker, Rykoff  \&
  Evrard}{Wechsler et~al.}{2021}]{Wechsler+21}
Wechsler R.~H.,  DeRose J.,  Busha M.~T.,  Becker M.~R.,  Rykoff E.,   Evrard
  A.,  2021, arXiv e-prints, p. arXiv:2105.12105

\bibitem[\protect\citeauthoryear{Weinberger et~al.,}{Weinberger
  et~al.}{2017}]{Weinberger+17}
Weinberger R.,  et~al., 2017, \mn@doi [Monthly Notices of the Royal
  Astronomical Society] {10.1093/mnras/stw2944}, 465, 3291–3308

\bibitem[\protect\citeauthoryear{Worthey}{Worthey}{1994}]{Worthey_1994}
Worthey G.,  1994, \mn@doi [The Astrophysical Journal Supplement Series]
  {10.1086/192096}, 95, 107

\bibitem[\protect\citeauthoryear{van~der Walt, Colbert  \& Varoquaux}{van~der
  Walt et~al.}{2011}]{numpy}
van~der Walt S.,  Colbert S.~C.,   Varoquaux G.,  2011, \mn@doi [Computing in
  Science Engineering] {10.1109/MCSE.2011.37}, 13, 22

\makeatother
\end{thebibliography}


\appendix

\section{Cluster richness vs mass}

\begin{figure*}
  \includegraphics [width=\linewidth] {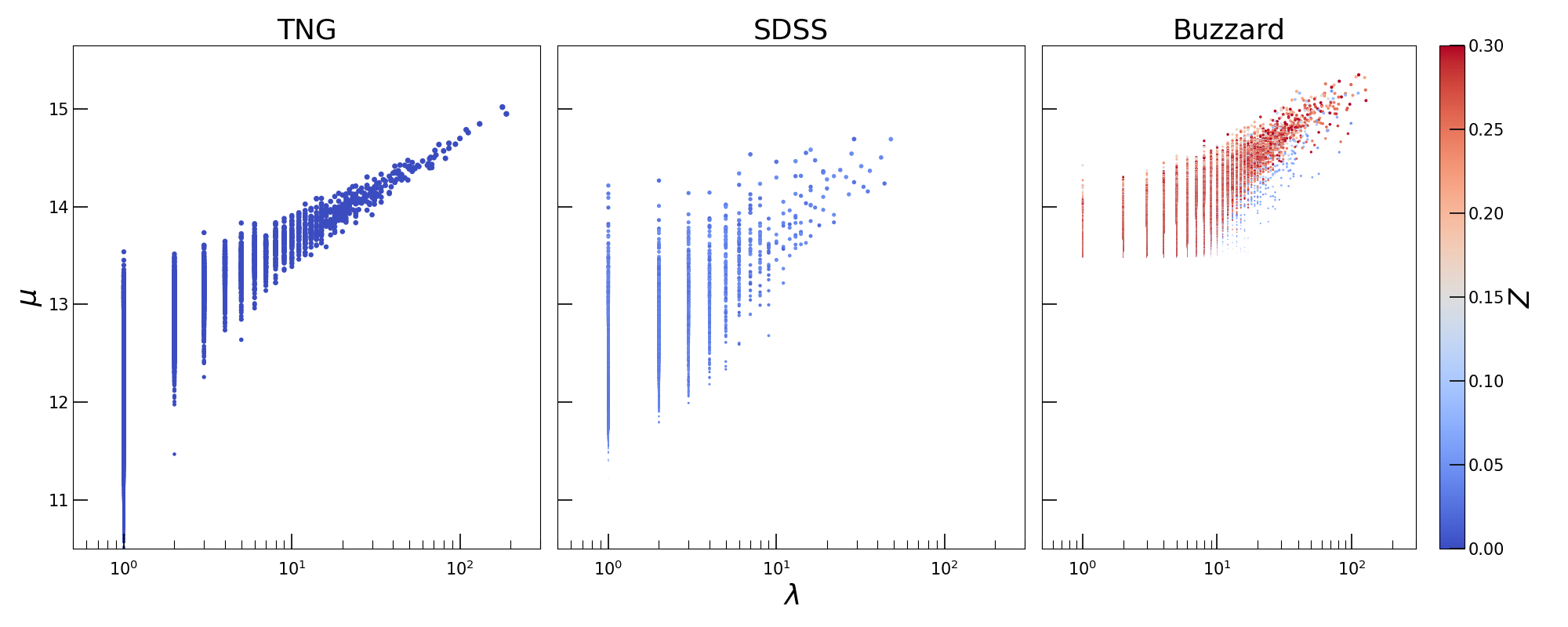}
  \caption{
    Mass--richness relation, colored by redshift. Note the cutoff in Buzzard data at $\mu = 13.5$.
  }
  \label{fig:richness_mass}
\end{figure*}

As noted in \citet{DeRose+19,DeRose+21}, Buzzard halos have fewer galaxies than observed, especially red galaxies at higher masses. 
Figure~\ref{fig:richness_mass} highlights this lack of red sequence galaxies. 
This figure also shows the lack of redshift evolution in SDSS clusters vs the significant redshift evolution in Buzzard clusters.

\newpage 
\section{Cluster counts by various thresholds} \label{apx:N_counts} 

\begin{figure}
  \includegraphics[width=\linewidth]{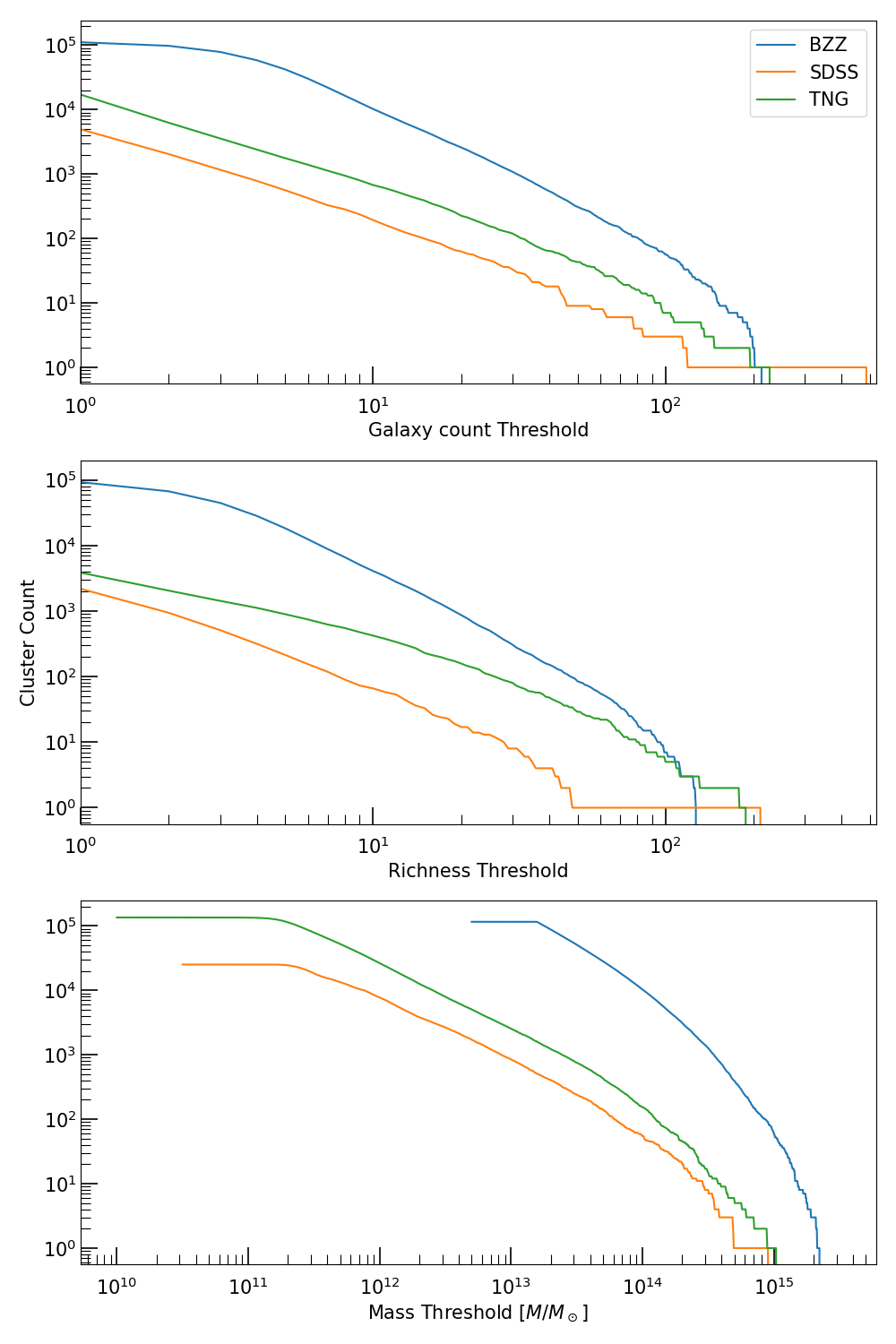}
  \caption{
    Number of clusters at given threshold, vis.: galaxy count $N_{\rm gal}$, richness $\lambda$, and mass $M$. 
  }
  \label{fig:count_vs_threshold}
\end{figure}

\begin{table}\centering
  \begin{tabular}{l r r r}
    \hline 
    & TNG & SDSS & Buzzard \\
    
    $N_{\rm tot}$ & 134953 & 25117 & 115508 \\
    \hline 
    $\lambda > 0$ & 18\% & 45\% & 95\% \\ 
    $N_{\rm gal} \geq 2$ & 12.6\% & 19.6\% & 95.9\% \\ 
    $N_{\rm gal} \geq 3$ & 4.6\% & 8.1\% & 84.3\% \\ 
    $N_{\rm gal} \geq 4$ & 2.6\% & 4.6\% & 67.8\% \\ 
    $\lambda \geq 10$ & 0.35\% & 0.29\%& 4.4\% \\
    $\lambda \geq 20$ & 0.13\% & 0.08\% & 0.9\% \\
    $\lambda \geq 30$ &  0.06\%& 0.03\% & 0.3\% \\
    \hline 
  \end{tabular}
  \caption{
    Fraction of clusters in each dataset above given thresholds. 
  }
  \label{tab:count_thresholds}
\end{table}

Table~\ref{tab:count_thresholds} shows percentages of each dataset which lie above given thresholds, complimentary to figure~\ref{fig:count_vs_threshold}. 
  The first threshold of $\lambda > 0$ relates to the capacity of red sequence based cluster finding algorithms such as redMaPPer to identify such clusters. 
  The second threshold of $N_{\rm gal} \geq 2$ relates to whether the cluster color reported is a summed color or just a single galaxy's color. 
  If $N_{\rm gal} \geq 3$, then satellite colors are averaged across galaxies, rather than being that of only a single satellite. 
  The threshold of $N_{\rm gal} \geq 4$ gives whether or not magnitude gap $M_{1,4}$ may be calculated for this cluster. 
The various richness thresholds (10, 20, 30) give more realistic cuts for observational data analysis.

\bsp	
\label{lastpage}
\end{document}